\documentclass[12pt,notitlepage,fleqn]{article}
\usepackage{graphicx}
\usepackage{amsmath}
\usepackage{indentfirst}
\usepackage{amsfonts}
\usepackage{amssymb}
\newtheorem{problem}{Hypothesis}

\begin{document}

\begin{center}
{\LARGE The Body Center Cubic Quark Lattice Model}

{\small (A Modification and Further Development of the Quark Model)}

{\small \bigskip}

{\normalsize Jiao-Lin Xu }

{\small The Center for Simulational Physics, The Department of Physics and Astronomy}

{\small The} {\small University of Georgia, Athens, GA 30602, U. S. A.}

E- mail: {\small jxu@hal.physast.uga.edu}

\ \ \ \ \ 

\ \ \ \ \ \ \ \ \ \ \ \ \ \ \ \ \ \ \ \ \ \ \ \ \ \ \ 

\textbf{Abstract}
\end{center}

{\small We assume that} {\small the u quarks and the d quarks constitute a
body center cubic quark lattice in the vacuum. Using energy band theory, we
deduce an excited quark spectrum (from the quark lattice). Using the
accompanying excitation concept, we deduce a baryon spectrum (including S, C,
b, I, Q, and mass) from the quark spectrum. With a phenomenological binding
energy formula, we deduce a meson spectrum (including S, C, b, I, Q, and mass)
from the quark spectrum. The baryon and meson spectra agree well with
experimental results. The BCC Quark Model predicts many new quarks
(u}$^{\prime}${\small (3), d}$^{\prime}${\small (6)), baryons (}$\Lambda^{0}%
${\small (4280), }$\Lambda_{C}^{+}${\small (6600), }$\Lambda_{b}^{0}%
${\small (9960)), and mesons (K(3597), D(5996), B(9504), }$\eta$%
{\small (5926), }$\Upsilon${\small (17805), T(1603) with I=2). The quarks
u}$^{\prime}${\small (3) and d}$^{\prime}${\small (6) and the meson T(1603)
have already been discovered.}

\section{Introduction}

The Quark Model \cite{QuarkModel} has successfully deduced intrinsic quantum
numbers ($I$, $S$, $C$, $b$, and $Q$) for all baryons and mesons. 1) It
cannot, however, deduce the mass spectrum of baryons or the mass spectrum of
mesons. 2) Some intrinsic quantum numbers and masses of the quarks are entered
``by hand'' \cite{Hand in}. 3) The Quark Model needs too many elementary
quarks. 4) It needs too many parameters \cite{Marshah}. 5) Confinement is a
very plausible idea, but to date its rigorous proof remains outstanding
\cite{Confinement}. 6) The large mass differences of the quarks (m$_{u}$ = 1 -
5 Mev, m$_{d}$ = 3 - 9 Mev, m$_{s}$ = 75 - 170 Mev, m$_{c}$ = 1150 - 1350 Mev,
m$_{b}$ = 4000 - 4400 Mev, m$_{t}$ = 174 Gev) \cite{Quark Mass(2000)} have
already broken down part of the mathematical foundation of the quark model
(the flavored SU(6), SU(5), might be SU(4) symmetries). 7) The quark masses of
the Quark Model are not large enough to build stable baryons and mesons. Thus,
the confinement assumption of the Quark Model cannot be maintained. Therefore,
the Quark Model needs modification and further development.

\ \ The BCC Quark Model (the Body Center Cubic Quark Lattice Model) \cite{BCC
Quark Model} is a good renovation of the Quark Model. It needs only 2
elementary quarks (u and d) and deduces an excited (from the vacuum) quark
spectrum (including S, C, b, I, Q, and M) \cite{Quark Spectrum}. Using the
quark spectrum, it can deduce the baryon spectrum \cite{Baryon Spectrum} and
the meson spectrum \cite{Net-Meson}, which both agree well with experimental
results \cite{Meson(2000)}. It also predicts some new quarks, baryons, and mesons{\small .}

\section{Fundamental Hypotheses}

From the Dirac sea concept{\small \ }and the Quark Model, there are infinite
u-quarks and d- quarks in the vacuum. There are super strong attractive forces
among the quarks. The forces make and hold the densest structure -- the body
center cubic quark lattice \cite{BCC MODEL}. Thus, we assume:

\begin{problem}
: There is only one family of elementary quarks in the vacuum state. They have
the baryon number $B=1/3$, spin $s=1/2$, isospin $I=1/2$, $S=C=b=0,$ and one
of the three color charges. The quarks with $I_{z}=+1/2$\ are called u quarks,
and the quarks with $I_{z}=-1/2$\ are called d quarks. There are super strong
attractive interactions among the quarks (colors). The forces make and hold an
infinite body center cubic (BCC) quark (u and d) lattice in the vacuum.
\end{problem}

\begin{problem}
: After a quark (q) is excited from the vacuum (q $\rightarrow$ q$^{\ast}$),
it can move in the vacuum space (the vacuum BCC quark lattice) freely; the
nearest quarks are accompanying excited simultaneously by the excited quark
q$^{\ast}$ as well. Since the super forces are short range and saturable (a 3
different colored quark system is a colorless one), there are only 2
accompanying excited quarks, u$^{\prime}$\ and d$^{\prime}$, for each excited
quark (q$^{\ast}$). The 2 quarks, u$^{\prime}$\ and d$^{\prime},$\ are just a
primitive cell of the BCC quark lattice. We call the excitation of the
primitive cell the accompanying excitation.
\end{problem}

The accompanying excited quark u$^{\prime}$ (or d$^{\prime}$) does not leave
its position in the quark lattice and cannot move freely in space since its
excited energy is not large enough. Their color charges, electric charges, and
baryon numbers, however, are temporarily excited from the vacuum state under
the influence of the excited quark q$^{\ast}$. Thus the excited energy of the
accompanying excited quarks (u$^{\prime}$\ and d$^{\prime}$) is much smaller
than the excited energy of the quark q$^{\ast}$. We have estimated that they
are ``very small and can be treated as a perturbation energy (0-order
approximation, m$_{u^{^{\prime}}}$ = m$_{d^{\prime}}$ = 0)'' in our earlier
paper \cite{Secsion II}. For convenience, we assume furthermore that
\begin{equation}
m_{u^{^{\prime}}}+m_{d^{\prime}}=\alpha M_{p}%
=938/137=7Mev,\label{Mass of Mu' and Md'}%
\end{equation}
where $M_{p}$ is the mass of proton, $\alpha$ =e$^{2}/c\hslash$ =1/137 -- the
fine structure constant. Comparing them with theoretical and experimental
results, we find that the masses of the quarks (u and d) of the Quark Model
(m$_{u}$ = 1 to 5 Mev, m$_{d}$ = 3 to 9 Mev) \cite{Quark Mass(2000)} are just
the quantities m$_{u^{\prime}}$ and m$_{d^{\prime}}$, and that the
accompanying excited quarks (u$^{\prime}$\ and d$^{\prime}$) have the same
intrinsic quantum numbers (B, S, s, I, I$_{Z}$ and Q) with the quarks (u and
d). Thus, the accompanying excited quark $u^{\prime}$ is the quark u
\cite{Quark Mass(2000)} with
\begin{equation}
B=1/3,\text{ }S=0,\text{ }s=I=1/2,\text{ }I_{z}=1/2,\text{ }Q=2/3,\text{
}m_{u^{\prime}}=3\text{ Mev;}\label{Quantum-0f-u}%
\end{equation}
and the accompanying excited quark $d^{\prime}$ is the quark d \cite{Quark
Mass(2000)} with
\begin{equation}
B=1/3,\text{ }S=0,s=I=1/2,\text{ }I_{z}=-1/2,\text{ }Q=-1/3,\text{
}m_{d^{\prime}}=6\text{ Mev}.\label{Quantum-0f-d}%
\end{equation}

The accompanying excitation is temporary for a cell (u$^{^{\prime}}$ and
d$^{^{\prime}}$). When the quark $q^{\ast}$ is excited from the vacuum, the
primitive cell (in which $q^{\ast}$ is located) undergoes an accompanying
excitation, which is due to $q^{\ast}$; but when $q^{\ast}$ leaves the cell,
the excitation of the cell disappears.\textbf{\ }Although the truly excited
cells are quickly changed -- one following another -- with the motion of
$q^{\ast}$, an accompanying excited primitive cell (u$^{\prime}$ and
d$^{\prime}$) always appears to accompany $q^{\ast},$ just as an electric
field always accompanies the original electric charge. \ 

\begin{problem}
:Due to the effect of the vacuum quark lattice, fluctuations of energy
$\varepsilon${\LARGE \ }and of intrinsic quantum numbers (such as the strange
number $S$) of an excited quark may exist. The fluctuation of the Strange
number, if it exists, is always $\Delta S=\pm1$ \cite{Real value of S}. From
the fluctuation of the Strange number, we will be able to deduce new quantum
numbers, such as Charmed number $C$ and Bottom number $b$.
\end{problem}

The excited quark $q^{\ast}$moving in space, in fact, is moving in a periodic
field of the body center cubic quark lattice. According to the energy band
theory \cite{Energy Band}, it will be in one of the energy bands [with a
degeneracy (deg) on a\textbf{\ }rotary fold\textbf{\ (}R) symmetry
axis\textbf{]} of the periodic field. Thus, it will have different quantum
numbers and energy (mass) when it is in the different energy band.

\begin{problem}
: The quantum numbers and masses of the excited quarks are determined as
follows (except for the 6 energy bands of the first Brillouin zone):
\end{problem}

\begin{enumerate}
\item  Baryon number $B$: When a quark is in a vacuum state, B = 0. If,
however, it is excited from the vacuum state, it has
\begin{equation}
B=1/3.\label{Baryon Number}%
\end{equation}

\item  Isospin number $I$: $I$\ is determined by the energy band degeneracy
$\deg$\ \cite{Energy Band}, where
\begin{equation}
\deg=2I+1.\label{IsoSpin}%
\end{equation}

\item  Strange number $S$: $S$\ is determined by the rotary fold $R$\ of the
symmetry axis \cite{Energy Band} with
\begin{equation}
S=R-4,\label{Strange Number}%
\end{equation}

where the number $4$\ is the highest possible rotary fold number of the BCC lattice.

\item  Electric charge $Q$: After obtaining $B,S$\ and $I$, we can find
$Q$\ from the Gell-Mann--Nishijiman relationship \cite{GellMann}\textbf{: }
\begin{equation}
Q=I_{z}+\frac{\text{1}}{2}(S+B).\label{Charge}%
\end{equation}

\item  Charmed number $C$\ and Bottom number $b$: If a degeneracy (deg) of an
energy band is smaller than the rotary fold R
\begin{equation}
\ \deg\ <\ R\ and\ R-\deg\ \neq2,\label{Condi-Sbar}%
\end{equation}
\ then formula (\ref{Strange Number}) will instead be
\begin{equation}
\bar{S}=R-4.\label{Strangebar}%
\end{equation}
\ From Hypothesis III ($\Delta S=\pm1$), the real value of $S$\ is
\begin{equation}
S=\bar{S}+\Delta S=S_{Axis}\pm1,\text{ \ \ \ \ \ if }\deg\ <\ R\text{ and
}R-d\neq2.\label{strangeflu}%
\end{equation}
The ``Strange number'', $S,$ \ in (\ref{strangeflu}) is not completely the
same as the Strange number in (\ref{Strange Number}). In order to compare it
with the experimental results, we would like to give it a new name under
certain circumstances. Based on Hypothesis III, the new names will be the
Charmed number and the Bottom number:
\begin{gather}
\text{if }S=+1,\text{ which originates from the fluctuation }\Delta
S=+1\text{, }\nonumber\\
\text{then we call it the Charmed number }C\text{ }(C=+1)\text{;}%
\label{Charmed}%
\end{gather}%
\begin{gather}
\text{if }S=-1,\text{ which originates from the fluctuation }\Delta
S=+1\text{, }\nonumber\\
\text{and if there is an energy fluctuation,}\nonumber\\
\text{then we call it the Bottom number }b\text{ }(b=-1)\text{.}\label{Bottom}%
\end{gather}
Similarly, we can obtain charmed strange quarks q$_{\Xi_{C}}^{\ast}$\ and
q$_{\Omega_{C}}^{\ast}$ \cite{Quark Spectrum}.

\item  We assume that the excited quark's static mass is the minimum energy of
the energy band that represents the excited quark:
\begin{equation}
m_{q^{\ast}}=\text{Minimum (}energy\text{) of the energy band}%
\label{Minimum(E)}%
\end{equation}

\item  The 6 energy bands of the first Brillouin zone represent the quark
q$_{N}^{\ast}(931)$ \cite{Quark Spectrum}.
\end{enumerate}

\begin{problem}
A meson is made of an excited quark $q_{i}^{\ast}$ and an excited antiquark
$\overline{q_{j}^{\ast}}$. The quantum numbers of the baryons and the mesons
are determined by sum laws from the constituent quarks:
\end{problem}%

\begin{equation}
\text{B=}\sum\text{B}_{q}\text{, S=}\sum\text{S}_{q}\text{, C=}\sum
\text{C}_{q}\text{, b=}\sum\text{b}_{q}\text{, Q=}\sum\text{Q}_{q}\text{,
}\overrightarrow{I}\text{=}\sum\overrightarrow{I_{q}}\text{.}\label{SUM-LAWS}%
\end{equation}

\section{The Spectrum of the Quarks}

When a excited quark is moving in the vacuum, it is moving, in fact, in the
periodic field ($V(\vec{r})$) of the vacuum quark lattice. Since the quark is
a Fermion, its motion equation should be the Dirac equation. According to the
renormalization theory \cite{Renormal}, the bare mass (m$_{q}$) of the quark
is much larger than the empirical values of the excited energies of the quark
in the energy bands. Thus, we use the Schr\"{o}dinger equation instead of the
Dirac equation (our results will show that this is a very good approximation)
\begin{equation}
\frac{\hslash^{2}}{2m_{q}}\nabla^{2}\Psi+(\varepsilon-V(\vec{r}))\Psi=0,
\end{equation}
where $V(\vec{r})$ denotes the strong interaction periodic field\ of the
vacuum quark lattice with body center cubic symmetries, and $m_{q}$ is the
bare mass of the elementary quark.

Using the energy band theory \cite{Energy Band} and the free particle
approximation (taking V($\vec{r}$) = V$_{0}$ constant and making the wave
functions satisfy the body center cubic periodic symmetries), we have
\begin{equation}
\frac{\hslash^{2}}{2m_{q}}\nabla^{2}\Psi+(\varepsilon-V_{0})\Psi
=0,\label{motion}%
\end{equation}
where $V_{0}$ is a constant potential. The solution of Eq. (\ref{motion}) is a
plane wave
\begin{equation}
\Psi_{\vec{k}}(\vec{r})=\exp\{-i(2\pi/a_{x})[(n_{1}-\xi)x+(n_{2}-\eta
)y+(n_{3}-\zeta)z]\},\label{wave}%
\end{equation}
where the wave vector $\vec{k}=(2\pi/a_{x})(\xi,\eta,\zeta)$, $a_{x}$ is the
periodic constant of the quark lattice, and $n_{1}$, $n_{2}$, and $n_{3}$ are
integers satisfying the condition n$_{1}$+n$_{2}$+n$_{3}$=$\pm$ even number or
$0.$ The wave functions must satisfy the symmetries of BCC periodic field.

The 0-order approximation of the energy (in \cite{Secsion II} , we assume
m$_{u^{^{\prime}}}$=m$_{d^{^{\prime}}}$=0$\rightarrow$V$_{0}$= 940 Mev; now we
find m$_{u^{^{\prime}}}$= 3 Mev, m$_{d^{^{\prime}}}$= 6 Mev$\rightarrow$
V$_{0}$= 931 Mev) is
\begin{equation}
\varepsilon^{(0)}\text{(}\vec{k}\text{,}\vec{n}\text{) = 931+360E(}\vec
{k}\text{,}\vec{n}\text{) Mev},\text{ E(}\vec{k}\text{,}\vec{n}\text{)=(n}%
_{1}\text{-}\xi\text{)}^{2}\text{+(n}_{2}\text{-}\eta\text{)}^{2}%
\text{+(n}_{3}\text{-}\zeta\text{)}^{2}\text{.}\label{mass}%
\end{equation}

Using the formulae (\ref{IsoSpin}), (\ref{Strange Number}), (\ref{Charge}),
(\ref{Charmed}), and (\ref{Bottom}) for quantum numbers and energy
(\ref{Minimum(E)}) of the quarks, we can find the quark spectrum \cite{Quark
Spectrum} :%

\begin{equation}%
\begin{tabular}
[c]{|l|l|l|l|l|l|l|l|l|l|l|l|l|l|}\hline
{\small q} & {\small B} & {\small S} & {\small C} & {\small b} & {\small I} &
\ \ \ \ \ \ {\small Q} & {\small q} & {\small B} & {\small S} & {\small C} &
{\small b} & {\small I} & {\small Q}\\\hline
{\small u} & {\small 0} & {\small 0} & {\small 0} & {\small 0} & {\small 0} &
{\small 0(vacuum state)} & {\small d} & {\small 0} & {\small 0} & {\small 0} &
{\small 0} & {\small 0} & {\small 0}\\\hline
{\small u}$^{^{\prime}}$ & {\small 1/3} & {\small 0} & {\small 0} & {\small 0}%
& {\small 1/2} & {\small 2/3} & {\small d}$^{^{\prime}}$ & {\small 1/3} &
{\small 0} & {\small 0} & {\small 0} & {\small 1/2} & {\small -1/3}\\\hline
{\small q}$_{N}^{\ast}$ & {\small 1/3} & {\small 0} & {\small 0} & {\small 0}%
& {\small 1/2} & {\small 2/3,-1/3} & {\small q}$_{S}^{\ast}$ & {\small 1/3} &
{\small -1} & {\small 0} & {\small 0} & {\small 0} & {\small -1/3}\\\hline
{\small q}$_{\Delta}^{\ast}$ & {\small 1/3} & {\small 0} & {\small 0} &
{\small 0} & {\small 3/2} & $\frac{5}{3}${\small ,}$\frac{2}{3}$%
{\small ,}$\frac{-1}{3}${\small ,}$\frac{-4}{3}$ & {\small q}$_{b}^{\ast}$ &
{\small 1/3} & {\small 0} & {\small 0} & {\small -1} & {\small 0} &
{\small -1/3}\\\hline
{\small q}$_{\Sigma}^{\ast}$ & {\small 1/3} & {\small -1} & {\small 0} &
{\small 0} & {\small 1} & {\small 2/3,-1/3,-4/3} & {\small q}$_{C}^{\ast}$ &
{\small 1/3} & {\small 0} & {\small 1} & {\small 0} & {\small 0} &
{\small 2/3}\\\hline
{\small q}$_{\Xi}^{\ast}$ & {\small 1/3} & {\small -2} & {\small 0} &
{\small 0} & {\small 1/2} & {\small -1/3,-4/3} & {\small q}$_{\Omega}^{\ast}$%
& {\small 1/3} & {\small -3} & {\small 0} & {\small 0} & {\small 0} &
{\small -4/3}\\\hline
{\small q}$_{\Sigma_{C}}^{\ast}$ & {\small 1/3} & {\small 0} & {\small 1} &
{\small 0} & {\small 1} & {\small 5/3,2/3,-1/3} & {\small q}$_{\Omega_{C}%
}^{\ast}$ & {\small 1/3} & {\small -2} & {\small 1} & {\small 0} & {\small 0}%
& {\small -1/3}\\\hline
{\small q}$_{\Xi_{C}}^{\ast}$ & {\small 1/3} & {\small -1} & {\small 1} &
{\small 0} & {\small 1/2} & {\small 2/3,-1/3} &  &  &  &  &  &  & \\\hline
\end{tabular}
\label{Quark Quantum Numbers}%
\end{equation}%

\begin{equation}%
\begin{tabular}
[c]{|l|}\hline
{\small \ The elementary\ quarks (u and d) in the BCC vacuum quark
lattice,\ m$_{u}$ = m$_{d}$ = 0.}\\\hline
\ \
\begin{tabular}
[c]{l}%
\ {\small The accompanying excited quarks u}$^{\prime}${\small and
d}$^{^{\prime}},${\small m}$_{u^{\prime}}$ {\small = 3 Mev, and m}%
$_{d^{\prime}}$ {\small = 6 Mev,}\\
{\small \ S=C=b=0; I=}$\frac{1}{2}$, I$_{z,u^{\prime}}$ = $\frac{1}{2}$,
I$_{z,d^{\prime}}$ = $\frac{-1}{2}$; {\small Q}$_{u^{\prime}}$ = $\frac{2}{3}
$, Q$_{d^{\prime}}$ = $\frac{-1}{3}$, spin s =$\frac{1}{2}$.
\end{tabular}
\\\hline
\ \ \ \ {\small The energy band excited states q}$^{\ast}${\small (from the
vacuum) of the\ quarks (u and d):}\\\hline
$%
\begin{tabular}
[c]{|l|l|l|l|}\hline
\textbf{q}$_{N}^{\ast}($\textbf{931)}$^{\#}
\begin{tabular}
[c]{l}%
{\small q}$_{u}^{\ast}${\small (931)}\\
{\small q}$_{d}^{\ast}${\small (931)}%
\end{tabular}
$ & \textbf{\ \ \ \ \ q}$_{S}^{\ast}($\textbf{1111)}$^{\#}$ & \textbf{q}%
$_{C}^{\ast}$\textbf{(2271)}$^{\#}$ & \textbf{q}$_{b}^{\ast}$\textbf{(5531)}%
$^{\#}$\\\hline
{\small \ q}$_{N}^{\ast}(${\small 1201),\ \ \ q}$_{\Delta}^{\ast}%
(${\small 1291)} & {\small \ q}$_{S}^{\ast}(${\small 1391), \ \ 2q}$_{\Xi
}^{\ast}(${\small 1291)} & {\small q}$_{C}^{\ast}${\small (2441)} &
{\small q}$_{b}^{\ast}${\small (9951)}\\\hline
{\small \ q}$_{N}^{\ast}(${\small 1471)}, {\small 2q}$_{\Delta}^{\ast}%
(${\small 1651)} & {\small \ q}$_{S}^{\ast}(${\small 2011), \ \ 3q}$_{\Xi
}^{\ast}(${\small 1471)} & {\small q}$_{C}^{\ast}${\small (2531)} &
{\small q}$_{b}^{\ast}${\small (15811)}\\\hline
{\small 2q}$_{N}^{\ast}(${\small 1831), \ q}$_{\Delta}^{\ast}(${\small 2011)}%
& {\small \ q}$_{S}^{\ast}(${\small 2451), \ \ 2q}$_{\Xi}^{\ast}%
(${\small 1651)\ } & {\small q}$_{C}^{\ast}${\small (2961)} & \\\hline
{\small 4q}$_{N}^{\ast}(${\small 1921), \ q}$_{\Delta}^{\ast}(${\small 2371)}%
& {\small \ q}$_{S}^{\ast}(${\small 2551), \ \ 2q}$_{\Xi}^{\ast}%
(${\small 1831)} & {\small q}$_{C}^{\ast}${\small (6591)} & {\small q}%
$_{\Xi_{c}}^{\ast}${\small (2541)}\\\hline
{\small 2q}$_{N}^{\ast}(${\small 2191), 4q}$_{\Delta}^{\ast}(${\small 2731)} &
{\small 3q}$_{S}^{\ast}(${\small 2641), \ \ q}$_{\Xi}^{\ast}(${\small 1921)} &
{\small q}$_{C}^{\ast}${\small (13791)} & {\small q}$_{\Xi_{c}}^{\ast}%
${\small (2631)}\\\hline
{\small 2q}$_{N}^{\ast}(${\small 2551), 3q}$_{\Delta}^{\ast}(${\small 3091)} &
{\small \ q}$_{S}^{\ast}(${\small 2731), \ \ 2q}$_{\Xi}^{\ast}($%
{\small 2011)\ \ } &  & {\small q}$_{\Xi_{c}}^{\ast}${\small (3161)}\\\hline
{\small 3q}$_{N}^{\ast}${\small (2641),} &
\ \ \ \ \ \ \ \ \ \ \ \ \ \ \ {\small 2q}$_{\Xi}^{\ast}(${\small 2191)} &
{\small q}$_{\Sigma_{C}}^{\ast}(${\small 2441)} & \\\hline
{\small 2q}$_{N}^{\ast}(${\small 2731)} & {\small \ q}$_{\Sigma}^{\ast}%
(${\small 1201), \ \ 2q}$_{\Xi}^{\ast}(${\small 2371)\ } & {\small q}%
$_{\Sigma_{C}}^{\ast}(${\small 2531)} & {\small q}$_{\Omega_{c}}^{\ast}%
${\small (2651)}\\\hline
& {\small 3q}$_{\Sigma}^{\ast}(${\small 1651), \ 3q}$_{\Xi}^{\ast}%
(${\small 2551)} & {\small q}$_{\Sigma_{C}}^{\ast}(${\small 2961)} &
{\small q}$_{\Omega_{c}}^{\ast}${\small (3471)}\\\hline
{\small \ } & {\small 5q}$_{\Sigma}^{\ast}(${\small 1921),\ \ 8q}$_{\Xi}%
^{\ast}(${\small 2731)} &  & \\\hline
{\small \ } & {\small 2q}$_{\Sigma}^{\ast}(${\small 2011)} &  & \\\hline
& {\small \ q}$_{\Sigma}^{\ast}(${\small 2371), \ \ q}$_{\Omega}^{\ast}%
${\small (1651)} &  & \\\hline
& {\small 3q}$_{\Sigma}^{\ast}(${\small 2551),\ \ q}$_{\Omega}^{\ast}%
${\small (2451)} &  & \\\hline
& {\small 2q}$_{\Sigma}^{\ast}(${\small 2641), \ q}$_{\Omega}^{\ast}%
${\small (3071)} &  & \\\hline
& {\small 3q}$_{\Sigma}^{\ast}(${\small 2731),\ \ q}$_{\Omega}^{\ast}%
${\small (3711)} &  & \\\hline
\end{tabular}
$\\\hline
\end{tabular}
\label{Quark-Spectrum}%
\end{equation}

\ \ \ \ \ \ \ \ \ \ \ \ \ \ \ \ \ \ 

\section{The Baryon Spectrum}

Since the baryon number of the system ($q^{\ast}u^{\prime}d^{\prime}$) is
\begin{equation}
B_{_{(}q^{\ast}u^{\prime}d^{\prime})}=B_{q^{\ast}}+B_{u^{\prime}}%
+B_{d^{\prime}}=1\label{B=1}%
\end{equation}
from (\ref{Quantum-0f-u}), (\ref{Quantum-0f-d}), (\ref{Baryon Number}), and
(\ref{SUM-LAWS}), the 3 quark system ($q^{\ast}u^{\prime}d^{\prime}$) is a
baryon. The mass of the baryon ($q^{\ast}u^{\prime}d^{\prime}$) equals
\begin{equation}
M_{_{(}q^{\ast}u^{\prime}d^{\prime})}=m_{q^{\ast}}+m_{u^{\prime}}%
+m_{d^{\prime}}.\label{baryon-Mass}%
\end{equation}
Using the sum laws (\ref{SUM-LAWS}) and (\ref{baryon-Mass}), from (\ref{Quark
Quantum Numbers}) and (\ref{Quark-Spectrum}), we can find the baryon spectrum
(including S, C, b, I, Q, and mass). We compare the theoretic results with the
experimental results \cite{Baryon(2000)} in Table 1 -- 4. In the comparison,
we do not take into account the angular momenta of the baryons. We assume that
the small differences of the masses in the same group of baryons (the same
kind of baryons with roughly the same masses but different angular momenta)
originate from their angular momenta. If we ignore this effect, their masses
should be the same. We use the average of the masses to represent the mass of
the group. We show quantum numbers of a baryon by its name.

\ \ \ \ \ \ \ \ \ \ \ \ \ \ \ \ \ \ \ \ \ \ \ \ \ \ %

\begin{tabular}
[c]{l}%
\ \ \ \ \ \ \ \ \ \ \ \ \ Table 1.\ The Ground States of the Baryons\\%
\begin{tabular}
[c]{|l|l|l||l|l|l|}\hline
{\small Theory} & {\small Experiment} & $\frac{\Delta\text{M}}{\text{M}}%
${\small \%} & {\small Theory} & {\small Experiment} & $\frac{\Delta\text{M}%
}{\text{M}}${\small \%}\\\hline
{\small N(940)} & {\small N(939)} & {\small 0.1 \%} & $\Omega^{-}%
${\small (1660)} & $\Omega^{-}${\small (1672)} & {\small 0.7 \%}\\\hline
$\Lambda${\small (1120)} & $\Lambda${\small (1116)} & {\small 0.4 \%} &
$\Lambda_{c}^{+}${\small (2280)} & $\Lambda_{c}^{+}${\small (2285)} &
{\small 0.2 \%}\\\hline
$\Sigma${\small (1210)} & $\Sigma${\small (1193)} & {\small 1.4 \%} &
$\Lambda_{b}^{0}${\small (5540)} & $\Lambda_{b}^{0}${\small (5624)} &
{\small 1.5 \%}\\\hline
$\Xi${\small (1300)} & $\Xi${\small (1318)} & {\small 1.4 \%} &  &  & \\\hline
\end{tabular}
\end{tabular}
\ \ \ \ \ \ \ \ \ \ \ \ \ \ \ \ \ \ \ \ \ \ \ \ \ 

\ \ \ \ \ \ \ \ \ \ \ \ \ \ \ \ \ \ \ \ 

{\small \ \ \ \ \ \ \ \ \ \ \ \ \ \ \ \ \ \ \ \ \ \ \ \ \ \ \ \ \ \ \ \ \ \ \ \ \ \ \ \ \ \ \ \ \ \ \ \ \ \ \ \ \ \ \ \ \ \ \ \ \ \ \ \ \ \ \ \ \ \ \ \ \ \ \ \ \ \ \ \ \ \ \ \ \ \ \ \ }%

\begin{tabular}
[c]{l}%
\ \ \ \ Table 2. The Unflavored Baryons $N$ and $\Delta$ ($S$= $C$=$b$ = 0)\\%
\begin{tabular}
[c]{|l|l|l||l|l|l|}\hline
Theory & Experiment & $\frac{\Delta\text{M}}{\text{M}}\%$ & Theory &
Experiment & $\frac{\Delta\text{M}}{\text{M}}\%$\\\hline
{\small \={N}(1255)} &  &  & $\bar{\Delta}${\small (1255)} & $\bar{\Delta}%
${\small (1232)} & {\small 1.9}\\\hline
{\small \={N}(1480)} & {\small \={N}(1498)} & {\small 1.2} &  &  & \\\hline
{\small \={N}(1660)} & {\small \={N}(1689)\ \ } & {\small 1.7} & $\bar{\Delta
}${\small (1660)} & $\bar{\Delta}${\small (1640)} & {\small 1.2}\\\hline
{\small \={N}(1915)} & {\small \={N}(1923)}$^{\ast}$ & {\small 3.9} &
$\bar{\Delta}${\small (1953)} & $\bar{\Delta}${\small (1923)} & {\small 1.6}%
\\\hline
{\small \={N}(2200)} & {\small \={N}(2220)} & {\small 0.9} &  &  & \\\hline
{\small N(2380)} & \ \ \ \ ? &  & $\Delta${\small (2380)} & $\Delta
${\small (2420)} & {\small 1.7}\\\hline
$\overline{\text{{\small N}}}${\small (2628)} & {\small N(2600)} &
{\small 1.1} & {\small 3}$\Delta${\small (2650)} & \ \ \ \ \ ? & \\\hline
{\small 6N(2740)} & {\small N(2700)}$^{\ast}$ & {\small 1.5} & {\small 4}%
$\Delta${\small (2750)} & $\Delta${\small (2740)}$^{\ast}$ & \\\hline
\end{tabular}
\end{tabular}

\ \ \ \ \ \ \ \ \ \ \ \ \ \ \ \ \ \ \ \ \ \ \ \ \ \ \ \ \ \ \ \ \ \ \ \ \ \ \ %

\begin{tabular}
[c]{l}%
\ \ \ \ \ \ \ \ \ \ \ Table 3. The Strange Baryons $\Lambda$ and $\Sigma$
($S=-1$)\\%
\begin{tabular}
[c]{|l|l|l||l|l|l|}\hline
Theory & Experiment & $\frac{\Delta\text{M}}{\text{M}}\%$ & Theory &
Experiment & $\frac{\Delta\text{M}}{\text{M}}\%$\\\hline
$\Lambda(1120)$ & $\Lambda(1116)$ & {\small 0.36\%} & $\Sigma(1210)$ &
$\Sigma(1193)$ & {\small 1.4 \%}\\\hline
$\overline{\mathbf{\Lambda}}(1400)$ & $\overline{\mathbf{\Lambda}}(1463)$ &
{\small 4.3 \%} & $\bar{\Sigma}(1350)$ & $\bar{\Sigma}(1385)$ & {\small 2.5
\%}\\\hline
$\bar{\Lambda}(1660)$ & $\bar{\Lambda}(1653)$ & {\small 0.4 \% } &
$\bar{\Sigma}(1660)$ & $\bar{\Sigma}(1714)$ & {\small 3.2 \%}\\\hline
$\bar{\Lambda}(1930)$ & $\bar{\Lambda}(1830)$ & {\small 5.5 \%} & $\bar
{\Sigma}(1930)$ & $\bar{\Sigma}(1928)$ & {\small .10 \%}\\\hline
$\bar{\Lambda}(2020)$ & $\Lambda(2105)$ & {\small 4.0 \%} & $\Sigma(2020)$ &
$\Sigma(2030)$ & {\small .50 \%}\\\hline
$\bar{\Lambda}(2420)$ & $\bar{\Lambda}(2350)$ & {\small 3.0 \%} & $\bar
{\Sigma}(2380)$ & $\bar{\Sigma}(2353)^{\#}$ & {\small 1.2 \%}\\\hline
$\Lambda(2560)$ & $\Lambda(2585)^{\ast}$ & {\small 1.0 \%} & $\Sigma
${\small (2650)} & $\Sigma(2620)$ & {\small 1.1\%}\\\hline
$\Lambda${\small (2650)} & \ \ \ \ \ ? &  & $\Sigma${\small (2740)} &
\ \ \ \ \ ? & \\\hline
\end{tabular}
\end{tabular}
\ \ \ \ \ \ \ \ \ \ \ \ \ \ \ \ \ \ \ \ \ \ \ \ \ \ \ \ \ \ \ \ \ \ \ \ \ \ \ \ \ \ \ \ \ \ \ \ \ \ \ \ \ \ \ \ \ \ \ \ \ \ \ \ \ \ \ \ \ \ \ \ \ \ \ \ \ \ \ \ \ \ \ \ \ \ \ \ \ \ \ \ \ \ \ 

\qquad{\small \# }$\bar{\Sigma}${\small (2353)}$^{\#}${\small =1/2(}$\Sigma
${\small (2250)+}$\Sigma${\small (2455))}

\ \ \ \ \ \ \ \ \ \ \ \ \ \ \ \ \ \ \ \ \ \ \ \ %

\begin{tabular}
[c]{l}%
\ \ \ \ \ \ \ \ \ \ \ \ \ Table 4. The $\Xi$, $\Omega$, $\Lambda_{c}^{+}$,
$\Sigma_{c}$, and $\Omega_{C}$ Baryons\\%
\begin{tabular}
[c]{|l|l|l||l|l|l|}\hline
Theory & Experim. & $\frac{\Delta\text{M}}{\text{M}}\%$ & Theory &
Experiment & $\frac{\Delta\text{M}}{\text{M}}\%$\\\hline
$\Xi(1300)$ & $\Xi(1318)$ & {\small 1.4 \%} & $\Omega(1660)$ & $\Omega(1672)$%
& {\small 0.7 \%}\\\hline
$\Xi(1480)$ & $\Xi(1530)$ & {\small 3.3 \% } & $\bar{\Omega}(2460)$ &
$\bar{\Omega}(2367)$ & {\small 3.9 \%}\\\hline
$\Xi(1660)$ & $\Xi(1690)$ & {\small 1.8 \%} & $\Omega(3080)$ & \ \ \ \ \ \ ? &
\\\hline
$\Xi(1840)$ & $\Xi(1820)$ & {\small 1.1 \%} &  &  & \\\hline
$\Xi(1930)$ & $\Xi(1950)$ & {\small 1.1 \%} & $\Lambda_{c}^{+}${\small (2280)}%
& $\Lambda_{c}^{+}${\small (2285)} & {\small 0.22 \%}\\\hline
$\Xi(2020)$ & $\Xi(2030)$ & {\small 1.0 \%} & $\bar{\Lambda}_{c}^{+}%
${\small (2540)} & $\bar{\Lambda}_{c}^{+}${\small (2609)} & {\small 2.3
\%}\\\hline
$\Xi(2200)$ & $\Xi(2250)^{\ast}$ & {\small 0.5 \%} & $\bar{\Lambda}_{c}^{+}%
${\small (2970)} & \ \ \ \ \ \ \ ? & \\\hline
$\Xi(2380)$ & $\Xi(2370)^{\ast}$ & {\small 0.4.2 \%} &  &  & \\\hline
$\Xi${\small (2560)} & \ \ \ \ \ ? &  & $\Lambda_{b}^{0}${\small (5540)} &
$\Lambda_{b}^{0}${\small (5624)} & {\small 1.5\%}\\\hline
$\Xi${\small (2740)} & \ \ \ \ \ ? &  & $\Lambda_{b}^{0}${\small (9960)} &
\ \ \ \ \ \ \ ? & \\\hline
&  &  &  &  & \\\hline
$\Xi_{C}${\small (2550)} & $\overline{\Xi_{C}}${\small (2523)} & {\small 1.1
\%} & $\overline{\Sigma_{c}}${\small (2495)} & $\overline{\Sigma_{c}}%
${\small (2488)} & {\small .28 \%}\\\hline
$\Xi_{C}${\small (2640)} & $\overline{\Xi_{C}}${\small (2730)} & {\small 3.2
\%} & $\Sigma_{c}${\small (2970)} & \ \ \ \ \ \ \ ? & \\\hline
$\Xi_{C}${\small (3170)} & \ \ \ \ \ ? &  &  &  & \\\hline
&  &  & $\Omega_{C}${\small (2660)} & $\Omega_{C}${\small (2704)} &
{\small 1.6 \%}\\\hline
&  &  & $\Omega_{C}${\small (3480)} & \ \ \ \ \ \ \ ? & \\\hline
\end{tabular}
\end{tabular}

In Table 1 -- Table 4, we see that the theoretical baryon spectrum agrees well
with the experimental results \cite{Baryon(2000)}.

\section{The Meson Spectrum}

According to the Quark Model, a meson is composed of a quark and an antiquark.
Using the quark quantum numbers (\ref{Quark Quantum Numbers}) and the sum law
(\ref{SUM-LAWS}), we can find the quantum numbers (S, C, b, I, Q) of the
mesons. Using a phenomenological formula of the meson binding energy
(\ref{E(i,j)}) and the mass of the quarks (\ref{Quark-Spectrum}), we can find
the masses [M(q$_{i}\overline{q}_{j}$)=m$_{q_{i}}$+m$_{\overline{q_{j}}}%
$+E$_{bind}$] of the mesons. The binding energy E$_{B}(i,j)$ \cite{Net-Meson}%
\ of a quark (q$_{i}^{\ast}$) and an antiquark ($\overline{q_{j}^{\ast}}$) in
the meson M(q$_{i}^{\ast}\overline{q_{j}^{\ast}}$) is\ \ \ \ \ %

\begin{equation}
\text{E}_{B}\text{(i,j)}=\text{-1723+}\text{100\{N+[2-}\left|  S\right|
\text{+}\delta_{ng}\text{f(I,S,C)](1-}\delta_{ij}\text{)+2}\delta
_{ng}\text{+(1-}\widetilde{m}\text{)-}\delta_{SP}N\text{\}+25A.}\label{E(i,j)}%
\end{equation}
Where N $=$ $\left|  S_{i\text{ }or\text{ j}}\right|  $+1.5$\left|  C_{i\text{
}or\text{ j}}\right|  $+3$\left|  b_{i\text{ }or\text{ j}}\right|  $, S is the
strange number, C is the charmed number, and b is the bottom number$.$ If
q$_{i}^{\ast}$ and $\overline{q_{j}^{\ast}}$ have the same S, C, or b,
N$_{i\text{ }or\text{ j}}$ = N$_{i}$. $\delta_{ng}$ = 1 if q$_{i}^{\ast}$ or
$\overline{q_{j}^{\ast}}$ is not ground state (or quark that is born from the
single band of the $\Delta$-axis and the $\Lambda$-axis). f(I,S,C) = -1.5
(SI-$\overline{SI}$)+$\Delta$I-$\delta$S-$2.5\left|  C\right|  $, SI$_{q}$ is
strange number S times ($\times$) isospin I of q$_{i}^{\ast};\overline{SI}$ is
strange number S times ($\times$) isospin I of $\overline{q_{j}^{\ast}}$.
$\Delta$I = $\left|  \text{I}_{i}\text{-I}_{\overline{j}}\right|  ,$ $\delta$S
= $\left|  \text{S}_{i}\text{-S}_{j}\right|  $, $\Delta$C = $\left|
\text{C}_{i}\text{-C}_{j}\right|  ,$ $\widetilde{m}=m_{q_{i}}m_{\overline
{q_{j}}}/m_{g}m_{\overline{g}}$. $\delta_{SP}$ = 1 if i = j and both quarks
(q$_{i}^{\ast}$ and $\overline{q_{i}^{\ast}})$ are born from the single binds;
$\delta_{SP}$=0 in all other cases. A = G$_{q}$-G$_{\overline{q}}$-SI$_{q}%
$+SI$_{\overline{q}}$, G = S+1.5C+3b. Although the above formula looks very
complex, we can simplify it into seven cases \cite{Net-Meson}; thus, it
becomes very simple and easy to use. For example, for ground quark pairs, from
(\ref{E(i,j)}) we have
\begin{equation}
\text{E}_{B}\text{(i, i) = -1723+100N}_{i}\text{+50(G}_{q}\text{-SI}%
_{q}\text{).}\label{Ground Pair}%
\end{equation}
Using (\ref{Ground Pair}), we find the mesons that are composed of the ground
quark pairs:
\begin{equation}%
\begin{tabular}
[c]{|l|l|}\hline
q$_{N}^{\ast}$({\small 931)}$\overline{\text{q}_{N}^{\ast}\text{{\small (931)}%
}}$=$\pi${\small (139)[}$\pi${\small (139)]} & q$_{C}^{\ast}${\small (2271)}%
$\overline{\text{{\small q}}_{C}^{\ast}\text{{\small (2271)}}}$={\small J/}%
$\Psi${\small (3044)[J/}$\Psi${\small (3097)]}\\\hline
q$_{S}^{\ast}${\small (1111)}$\overline{\text{{\small q}}_{S}^{\ast
}\text{{\small (1111)}}}$=$\eta${\small (549)[}$\eta${\small (547)]} &
q$_{b}^{\ast}${\small (5531)}$\overline{q_{b}^{\ast}(5531)}$=$\Upsilon
${\small (9489) [}$\Upsilon${\small (9460)].}\\\hline
\end{tabular}
\label{Mesons of Ground Pairs}%
\end{equation}
In the fashion similarly to (\ref{Ground Pair}) and (\ref{Mesons of Ground
Pairs}), we can find the meson spectrum \cite{Net-Meson}.

We compare the theoretical results with the experimental results using Table 5
-- 7. In the comparison, as is the case with the baryon spectrum, we do not
take into account the angular momenta but show the quantum numbers of a meson
with its name.

\ Table 5 \ Light Unflavored Mesons (S=C=b=0) and Bottom Mesons (b=$\pm1)$%

\begin{tabular}
[c]{|l|}\hline
\ \ \ \ \ \ \ \ {\small Light Unflavored Mesons} {\small (I = 0)
}\ \ \ \ \ \ \ \ \ \ \ \ \ \ \ {\small Light Unflavored Mesons} {\small (I =
1)}\\\hline%
\begin{tabular}
[c]{|l|l|}\hline
{\small \ \ \ \ \ The\ BCC Quark Model} & {\small Exper.}\\\hline
\ q$_{i}$(m$_{k}$)$\overline{q_{j}(m_{l})}=\eta(m)$ & \ M(m)\\\hline
{\small q}$_{S}^{\ast}${\small (1111)}$\overline{q_{S}^{\ast}%
\text{{\small (1111)}}}${\small =}$\eta${\small (549)} & $\eta${\small (547)}%
\\\hline
{\small q}$_{N}^{\ast}${\small (931)}$\overline{q_{N}^{\ast}%
\text{{\small (1201)}}}${\small =}$\eta${\small (780)} & $\omega
${\small (782)}\\\hline
{\small q}$_{N}^{\ast}${\small (1201)}$\overline{q_{N}^{\ast}%
\text{{\small (1201)}}}${\small =}$\eta${\small (813)} & {\small f}$_{0}%
${\small (800)}\\\hline
{\small q}$_{S}^{\ast}${\small (1391)}$\overline{q_{S}^{\ast}%
\text{{\small (1391)}}}${\small =}$\eta${\small (952)} & $\eta^{\prime}%
${\small (958)}\\\hline
{\small q}$_{\Delta}^{\ast}${\small (1291)}$\overline{q_{\Delta}^{\ast}%
(1291)}${\small =}$\eta${\small (967)} & {\small f}$_{0}${\small (980)}%
\\\hline
{\small q}$_{N}^{\ast}${\small (931)}$\overline{q_{N}^{\ast}%
\text{{\small (1471)}}}${\small =}$\eta${\small (1021)} & $\phi$%
{\small (1020)}\\\hline
q$_{S}^{\ast}${\small (1111)}$\overline{q_{S}^{\ast}\text{{\small (1391)}}}
${\small =}$\eta${\small (1204)} & {\small h}$_{1}${\small (1170)}\\\hline
{\small q}$_{N}^{\ast}${\small (1471)}$\overline{q_{N}^{\ast}%
\text{{\small (1471)}}}${\small =}$\eta${\small (1269)} & $\overline{\eta}%
${\small (1283)}\\\hline
{\small q}$_{N}^{\ast}${\small (931)}$\overline{q_{N}^{\ast}%
\text{{\small (1831)}}}${\small =}$\eta${\small (1342)} & $\overline{\eta}%
${\small (1375)}\\\hline
{\small q}$_{N}^{\ast}${\small (931)}$\overline{q_{N}^{\ast}%
\text{{\small (1921)}}}${\small =}$\eta${\small (1422)} & $\overline{\eta}%
${\small (1428)}\\\hline
{\small q}$_{\Delta}^{\ast}${\small (1651)}$\overline{q_{\Delta}^{\ast
}\text{{\small (1651)}}}${\small =}$\eta${\small (1565)} & $\overline{\eta}%
${\small (1525)}\\\hline
{\small q}$_{N}^{\ast}${\small (931)}$\overline{q_{N}^{\ast}%
\text{{\small (2191)}}}${\small =}$\eta${\small (1664)} & $\overline{\eta}%
${\small (1656)}\\\hline
{\small q}$_{S}^{\ast}${\small (1111)}$\overline{q_{S}^{\ast}%
\text{{\small (2011)}}}${\small =}$\eta${\small (1768)} & $\overline{\eta}%
${\small (1735)}\\\hline
{\small q}$_{N}^{\ast}${\small (1831)}$\overline{q_{N}^{\ast}%
\text{{\small (1831)}}}${\small =}$\eta${\small (1852)} & $\overline{\eta}%
${\small (1850)}\\\hline
{\small q}$_{N}^{\ast}${\small (931)}$\overline{q_{N}^{\ast}%
\text{{\small (2551)}}}${\small =}$\eta${\small (1985)} & $\overline{\eta}%
${\small (1930)}\\\hline
{\small q}$_{N}^{\ast}${\small (931)}$\overline{q_{N}^{\ast}%
\text{{\small (2641)}}}${\small =}$\eta${\small (2065)} & $\overline{\eta}%
${\small (2030)}\\\hline
{\small q}$_{N}^{\ast}${\small (931)}$\overline{q_{N}^{\ast}%
\text{{\small (2731)}}}${\small =}$\eta${\small (2146)} & $\overline{\eta}%
${\small (2199)}\\\hline
{\small q}$_{S}^{\ast}${\small (1111)}$\overline{q_{S}^{\ast}%
\text{{\small (2641)}}}${\small =}$\eta${\small (2341)} & $\overline{\eta}%
${\small (2313)}\\\hline
{\small q}$_{S}^{\ast}${\small (1111)}$\overline{q_{S}^{\ast}%
\text{{\small (2731)}}}${\small =}$\eta${\small (2423)} & $\eta$%
{\small (2510)}\\\hline
{\small q}$_{\Delta}^{\ast}${\small (2731)}$\overline{q_{\Delta}^{\ast
}\text{{\small (2731)}}}${\small =}$\eta${\small (3178)} & $\chi
${\small (3250)}\\\hline
{\small q}$_{\Sigma}^{\ast}${\small (2641)}$\overline{q_{\Sigma}^{\ast
}\text{{\small (2641)}}}${\small =}$\eta${\small (3479)} & \ \ \ \ ?\\\hline
{\small q}$_{S}^{\ast}${\small (4271)}$\overline{q_{S}^{\ast}%
\text{{\small (4271)}}}${\small =}$\eta${\small (5926)} & \ \ \ \ ?\\\hline
& \\\hline
& \\\hline
\end{tabular}%
\begin{tabular}
[c]{|l|l|}\hline
{\small \ \ The BCC Quark Model} & {\small Exper.}\\\hline
\ q$_{i}$(m$_{k}$)$\overline{q_{j}(m_{l})}=\pi(m)$ & {\small M(m)}\\\hline
{\small q}$_{N}^{\ast}${\small (931)}$\overline{q_{N}^{\ast}(931)}$%
{\small =}$\pi${\small (139)} & $\pi${\small (139)}\\\hline
{\small q}$_{N}^{\ast}${\small (931)}$\overline{q_{N}^{\ast}%
\text{{\small (1201)}}}${\small =}$\pi${\small (780)} & $\rho${\small (770)}%
\\\hline
{\small q}$_{N}^{\ast}${\small (931)}$\overline{q_{\Delta}^{\ast}(1291)}%
${\small =}$\pi${\small (960)} & {\small a}$_{0}${\small (980)}\\\hline
{\small q}$_{N}^{\ast}${\small (931)}$\overline{q_{\Delta}^{\ast
}\text{{\small (1651)}}}${\small =}$\pi${\small (1282)} & $\overline{\pi}%
${\small (1248)}\\\hline
{\small q}$_{N}^{\ast}${\small (931)}$\overline{q_{N}^{\ast}%
\text{{\small (1831)}}}${\small =}$\pi${\small (1342)} & $\overline{\pi}%
${\small (1340)}\\\hline
{\small q}$_{N}^{\ast}${\small (931)}$\overline{q_{N}^{\ast}%
\text{{\small (1921)}}}${\small =}$\pi${\small (1423)} & $\overline{\pi}%
${\small (1450)}\\\hline
{\small q}$_{N}^{\ast}${\small (931)}$\overline{q_{\Delta}^{\ast
}\text{{\small (2011)}}}${\small =}$\pi${\small (1603)} & $\pi${\small (1600)}%
\\\hline
{\small q}$_{N}^{\ast}${\small (931)}$\overline{q_{N}^{\ast}%
\text{{\small (2191)}}}${\small =}$\pi${\small (1664)} & $\overline{\pi}%
${\small (1687)}\\\hline
{\small q}$_{S}^{\ast}${\small (1111)}$\overline{q_{\Sigma}^{\ast
}\text{{\small (1921)}}}${\small =}$\pi${\small (1861)} & $\pi${\small (1800)}%
\\\hline
{\small q}$_{N}^{\ast}${\small (931)}$\overline{q_{\Delta}^{\ast
}\text{{\small (2371)}}}${\small =}$\pi${\small (1924)} & $\chi$%
{\small (2000)}\\\hline
{\small q}$_{N}^{\ast}${\small (931)}$\overline{q_{N}^{\ast}%
\text{{\small (2641)}}}${\small =}$\pi${\small (2065)} & $\pi${\small (2040)}%
\\\hline
{\small q}$_{N}^{\ast}${\small (931)}$\overline{q_{N}^{\ast}%
\text{{\small (2731)}}}${\small =}$\pi${\small (2146)} & $\overline{\pi}%
${\small (2125)}\\\hline
{\small q}$_{N}^{\ast}${\small (931)}$\overline{q_{\Delta}^{\ast
}\text{{\small (2731)}}}${\small =}$\pi${\small (2244)} & $\pi${\small (2250)}%
\\\hline
{\small q}$_{S}^{\ast}${\small (1111)}$\overline{q_{\Sigma}^{\ast
}\text{{\small (2551)}}}${\small =}$\pi${\small (2434)} & $\overline{\pi}%
${\small (2400)}\\\hline
& \\\hline
\ Bottom ({\small b=}$\pm1)$ Meson & {\small Exper.}\\\hline
$\overline{q_{b}^{\ast}\text{{\small (5531)}}}$q$_{N}^{\ast}$%
(931){\small =B(5164)} & {\small B(5279)}\\\hline
$\overline{q_{b}^{\ast}\text{{\small (5531)}}}q_{N}^{\ast}$({\small 1201}%
){\small =B(5655)} & {\small B(5732)}\\\hline
$\overline{q_{b}^{\ast}\text{{\small (5531)}}}q_{N}^{\ast}$({\small 1471}%
){\small =B(5896)} & \ \ \ \ \ ?\\\hline
\ Bottom Strange(S=$\pm1)$ & \\\hline
$\overline{q_{b}^{\ast}\text{{\small (5531)}}}$q$_{S}^{\ast}${\small (1111)=B}%
$_{S}${\small (5319)} & {\small B}$_{S}${\small (5370)}\\\hline
$\overline{q_{b}^{\ast}\text{{\small (5531)}}}q_{S}^{\ast}$({\small 1391}%
)={\small B}$_{S}${\small (5674)} & {\small B}$_{S}${\small (5850)}\\\hline
Bottom Charmed(C=-b=$1)$ & \\\hline
$\overline{q_{b}^{\ast}\text{{\small (5531)}}}q_{C}^{\ast}$({\small 2271}%
)={\small B}$_{C}${\small (6691)} & {\small B}$_{C}${\small (6400)}\\\hline
\end{tabular}
\\\hline
\end{tabular}

\qquad%

\begin{tabular}
[c]{l}%
\ \ \ \ \ \ \ \ \ \ \ \ Table 6 \ \ K Mesons, D Mesons, and D$_{S}$ Mesons\\%
\begin{tabular}
[c]{|l|l|}\hline
{\small \ \ \ \ \ The BCC Quark Model} & {\small Exper.}\\\hline
{\small q}$_{i}${\small (m}$_{k}${\small )}$\overline{q_{j}(m_{l})}%
=${\small K(m)} & {\small \ K(m)}\\\hline
q$_{N}^{\ast}$(931)$\overline{q_{S}^{\ast}\text{{\small (1111)}}}%
${\small =K(494)} & {\small K(494)}\\\hline
q$_{N}^{\ast}$(931)$\overline{\text{q}_{S}^{\ast}\text{{\small (1391)}}}%
${\small =K(899)} & {\small K(892)}\\\hline
q$_{N}^{\ast}$(931)$\overline{\text{q}_{\Sigma}^{\ast}\text{{\small (1651)}}}%
${\small =K(1310)} & {\small K(1270)}\\\hline
q$_{N}^{\ast}$(931)$\overline{\text{q}_{S}^{\ast}\text{{\small (2011)}}}%
${\small =K(1463)} & $\overline{K}${\small (1423)}\\\hline
q$_{N}^{\ast}$(931)$\overline{\text{q}_{\Sigma}^{\ast}\text{{\small (1921)}}}%
${\small =K(1556)} & {\small K(1580)}\\\hline
q$_{N}^{\ast}$(931)$\overline{\text{q}_{\Sigma}^{\ast}\text{{\small (2011)}}}%
${\small =K(1638)} & {\small K(1680)}\\\hline
{\small q}$_{N}^{\ast}${\small (2191)}$\overline{q_{S}^{\ast}%
\text{{\small (1111)}}}${\small =K(1769)} & $\overline{K}${\small (1775)}%
\\\hline
q$_{N}^{\ast}$(931)$\overline{\text{q}_{S}^{\ast}\text{{\small (2551)}}}%
${\small =K(1804)} & $\overline{K}${\small (1820)}\\\hline
q$_{N}^{\ast}$(931)$\overline{\text{q}_{\Sigma}^{\ast}\text{{\small (2371)}}}%
${\small =K(1966)} & $\overline{K}${\small (1965)}\\\hline
q$_{N}^{\ast}$(931)$\overline{\text{q}_{S}^{\ast}\text{{\small (2641)}}}%
${\small =K(2036)} & {\small K(2045)}\\\hline
q$_{N}^{\ast}$(931)$\overline{\text{q}_{\Sigma}^{\ast}\text{{\small (2641)}}}%
${\small =K(2211)} & {\small K(2250)}\\\hline
q$_{N}^{\ast}$(931)$\overline{\text{q}_{\Sigma}^{\ast}\text{{\small (2731)}}}%
${\small =K(2293)} & $\overline{K}${\small (2350)}\\\hline
q$_{N}^{\ast}$(931)$\overline{\text{q}_{\Sigma}^{\ast}\text{{\small (3091)}}}%
${\small =K(2621)} & {\small K(2500)}\\\hline
q$_{N}^{\ast}$(931)$\overline{\text{q}_{S}^{\ast}\text{{\small (4271)}}}%
${\small =K(3597)} & {\small \ \ \ \ \ \ ?}\\\hline
\end{tabular}%
\begin{tabular}
[c]{|l|l|}\hline
{\small \ \ The BCC Quark Model} & {\small \ Exper.}\\\hline
{\small \ \ \ \ q}$_{i}${\small (m}$_{k}${\small )}$\overline{q_{j}(m_{l})}%
${\small \ = D(m)} & {\small M(m)}\\\hline
{\small q}$_{C}^{\ast}${\small (2271)}$\overline{q_{N}^{\ast}%
\text{{\small (931)}}}${\small \ = D(1866)} & {\small D(1869)}\\\hline
{\small q}$_{C}^{\ast}${\small (2441)}$\overline{q_{N}^{\ast}%
\text{{\small (931)}}}${\small \ = D(2029)} & {\small D(2010)}\\\hline
{\small q}$_{C}^{\ast}${\small (2531)}$\overline{q_{N}^{\ast}%
\text{{\small (931)}}}${\small \ = D(2115)} & \ \ \ \ \ \ {\small ?}\\\hline
{\small q}$_{C}^{\ast}${\small (2271)}$\overline{q_{N}^{\ast}%
\text{{\small (1471)}}}${\small =D(2349)} & {\small D}$_{1}${\small (2420)}%
\\\hline
{\small q}$_{C}^{\ast}${\small (2961)}$\overline{q_{N}^{\ast}%
\text{{\small (931)}}}${\small \ = D(2526)} & {\small D(2507)}\\\hline
{\small q}$_{C}^{\ast}${\small (6591)}$\overline{q_{N}^{\ast}%
\text{{\small (931)}}}${\small \ = D(5996)} & \ \ \ \ \ \ {\small ?}\\\hline
& \\\hline
{\small \ \ \ \ q}$_{i}${\small (m}$_{k}${\small )}$\overline{q_{j}%
\text{(m}_{l}\text{)}}${\small \ = D}$_{S}${\small (m)} & \\\hline
{\small q}$_{C}^{\ast}${\small (2271)}$\overline{q_{S}^{\ast}%
\text{{\small (1111)}}}${\small \ = D}$_{S}${\small (2021)} & {\small D}$_{S}%
${\small (1969)}\\\hline
{\small q}$_{C}^{\ast}${\small (2271)}$\overline{q_{S}^{\ast}%
\text{{\small (1391)}}}${\small \ = D}$_{S}${\small (2126)} & {\small D}%
$_{S}^{\ast}${\small (2112)}\\\hline
{\small q}$_{C}^{\ast}${\small (2961)}$\overline{q_{S}^{\ast}%
\text{{\small (1111)}}\ }${\small = D}$_{S}${\small (2531)} & {\small D}$_{S}%
${\small (2555)}$^{\pm}$\\\hline
{\small q}$_{C}^{\ast}${\small (2271)}$\overline{q_{S}^{\ast}%
\text{{\small (2551)}}}${\small \ = D}$_{S}${\small (3333)} &
\ \ \ \ \ \ {\small ?}\\\hline
{\small q}$_{C}^{\ast}${\small (6591)}$\overline{q_{S}^{\ast}%
\text{{\small (1111)}}}${\small \ = D}$_{S}${\small (6151)} &
\ \ \ \ \ \ {\small ?}\\\hline
& \\\hline
\end{tabular}
\end{tabular}
\ \ \ \ \ \ \ \ 

\ \ \ \ \ \ \ \ \ \ \ \ \ \ \ \ \ \ \ \ \ \ \ \ \ \ \ \ \ \ \ \ \ \ \ \ \ \ \ \ \ \ \ \ \ \ \ \ %

\begin{tabular}
[c]{l}%
\ \ \ \ \ \ \ \ \ \ \ \ \ \ \ \ Table 7 Charmed Pair Mesons and Bottomed Pair
Mesons\ \\%
\begin{tabular}
[c]{|l|l|}\hline
\ \ \ \ The BCC Quark Model & Exper.\\\hline
\ \ \ \ \ \ q$_{i}$(m$_{k}$)$\overline{q_{j}(m_{l})}=M(m)$ & M(m)\\\hline
{\small q}$_{S}^{\ast}${\small (2551)}$\overline{q_{S}^{\ast}%
\text{{\small (2551)}}}${\small =}$\eta${\small (2902)} & $\eta_{C}%
${\small (2980)}\\\hline
{\small q}$_{C}^{\ast}${\small (2271)}$\overline{\text{q}_{C}^{\ast
}\text{{\small (2271)}}}${\small =J/}$\Psi${\small (3044)} & {\small J/}$\Psi
${\small (3097)}\\\hline
{\small q}$_{\Sigma}^{\ast}${\small (2641)}$\overline{q_{\Sigma}^{\ast
}\text{{\small (2641)}}}${\small =}$\eta${\small (3394)} & $\chi
${\small (3415)}\\\hline
{\small q}$_{\Sigma}^{\ast}${\small (2731)}$\overline{q_{\Sigma}^{\ast
}\text{{\small (2731)}}}${\small =}$\eta${\small (3535)} & $\chi
${\small (3510)}\\\hline
{\small q}$_{C}^{\ast}${\small (2441)}$\overline{q_{C}^{\ast}%
\text{{\small (2441)}}}${\small =}$\psi${\small (3568)} & $\chi$%
{\small (3556)}\\\hline
{\small q}$_{C}^{\ast}${\small (2271)}$\overline{q_{C}^{\ast}%
\text{{\small (2531)}}}${\small =}$\psi${\small (3693)} & $\psi$%
{\small (3686)}\\\hline
{\small q}$_{C}^{\ast}${\small (2531)}$\overline{q_{C}^{\ast}%
\text{{\small (2531)}}}${\small =}$\psi${\small (3740)} & $\psi$%
{\small (3770)}\\\hline
{\small q}$_{S}^{\ast}${\small (1111)}$\overline{q_{S}^{\ast}%
\text{{\small (4271)}}}${\small =}$\eta${\small (3952)} & $\psi$%
{\small (4040)}\\\hline
{\small q}$_{C}^{\ast}${\small (2271)}$\overline{\text{q}_{C}^{\ast
}\text{{\small (2961)}}}${\small =}$\psi${\small (4104)} & $\psi
${\small (4160)}\\\hline
{\small q}$_{C}^{\ast}${\small (2961)}$\overline{q_{C}^{\ast}%
\text{{\small (2961)}}}${\small =}$\psi${\small (4554)} & $\psi$%
{\small (4415)}\\\hline
q$_{C}^{\ast}${\small (6591)}$\overline{q_{C}^{\ast}\text{{\small (2271)}}}
${\small =}$\psi${\small (7374)} & \ \ \ \ \ {\small ?}\\\hline
\end{tabular}%
\begin{tabular}
[c]{|l|l|}\hline
\ \ \ \ The BCC Quark Model & \ \ Exper.\\\hline
\ \ \ \ \ q$_{i}$(m$_{k}$)$\overline{q_{j}(m_{l})}=M(m)$ & \ \ M(m)\\\hline
{\small q}$_{b}^{\ast}${\small (5531)}$\overline{q_{b}^{\ast}%
\text{{\small (5531)}}}${\small =}$\Upsilon${\small (9489)} & $\Upsilon
${\small (9460)}\\\hline
{\small q}$_{S}^{\ast}${\small (1111)}$\overline{q_{S}^{\ast}%
\text{{\small (10031)}}}${\small =}$\eta${\small (9734)} & $\chi
${\small (9888)}\\\hline
{\small q}$_{S}^{\ast}${\small (1391)}$\overline{q_{S}^{\ast}%
\text{{\small (10031)}}}${\small =}$\eta${\small (9955)} & $\Upsilon
${\small (10023)}\\\hline
{\small q}$_{S}^{\ast}${\small (2011)}$\overline{q_{S}^{\ast}%
\text{{\small (10031)}}}${\small =}$\eta${\small (10443)} & $\chi
${\small (10252)}\\\hline
{\small q}$_{C}^{\ast}${\small (6591)}$\overline{q_{C}^{\ast}%
\text{{\small (6591)}}}${\small =}$\psi${\small (10792)} & $\Upsilon
${\small (10355)}\\\hline
{\small q}$_{S}^{\ast}${\small (2451)}$\overline{q_{S}^{\ast}%
\text{{\small (10031)}}}${\small =}$\eta${\small (10791)} & $\Upsilon
${\small (10580)}\\\hline
{\small q}$_{S}^{\ast}${\small (2551)}$\overline{q_{S}^{\ast}%
\text{{\small (10031)}}}${\small =}$\eta${\small (10870)} & $\Upsilon
${\small (10860)}\\\hline
{\small q}$_{S}^{\ast}${\small (2641)}$\overline{q_{S}^{\ast}%
\text{{\small (10031)}}}${\small =}$\eta${\small (10941)} &
\ \ \ \ \ {\small ?}\\\hline
{\small q}$_{S}^{\ast}${\small (2731)}$\overline{q_{S}^{\ast}%
\text{{\small (10031)}}}${\small =}$\eta${\small (11012)} & $\Upsilon
${\small (11020)}\\\hline
{\small q}$_{b}^{\ast}${\small (5531)}$\overline{q_{b}^{\ast}(9951)}%
${\small =}$\Upsilon${\small (14329)} & \ \ \ \ \ {\small ?}\\\hline
{\small q}$_{b}^{\ast}${\small (9951)}$\overline{q_{b}^{\ast}(9951)}%
${\small =}$\Upsilon${\small (17805)} & \ \ \ \ \ {\small ?}\\\hline
\end{tabular}
\end{tabular}
\ \ \ \ \ \ \ \ \ \ \ \ \ \ \ \ \ \ \ \ \ \ \ \ \ \ \ \ \ \ \ \ \ \ \ \ \ \ \ \ \ \ \ \ \ \ \ \ \ \ \ \ \ \ \ \ \ \ \ \ \ \ \ \ \ \ \ \ \ \ \ \ \ \ \ \ \ \ \ \ \ \ \ \ \ \ \ \ \ \ \ \ \ \ \ \ \ \ \ \ \ \ \ \ \ \ \ \ \ \ \ \ \ \ \ \ \ \ \ \ \ \ \ \ \ \ \ \ \ \ \ \ \ \ \ \ \ \ \ \ \ \ \ \ \ \ \ \ \ \ \ \ \ \ \ \ \ \ \ \ \ \ \ \ \ \ \ \ \ \ \ \ \ \ \ \ \ \ \ \ \ \ \ \ \ \ \ \ \ \ \ \ \ \ \ \ \ \ \ \ \ \ \ \ \ \ \ \ \ \ \ \ \ \ \ \ \ \ \ \ \ \ \ \ \ \ \ \ \ \ \ \ \ \ \ \ \ \ \ \ \ \ \ \ \ \ \ \ \ \ \ \ \ \ \ \ \ \ \ \ \ \ \ \ \ \ \ \ \ \ \ \ \ \ \ \ \ \ \ \ \ \ \ \ \ \ \ \ \ \ \ \ \ \ \ \ \ \ \ \ \ \ \ \ \ \ \ \ \ \qquad
\ \ \ \ \ \ \ \ \ \ \ \ \ \ \ \ \ \ \ \ \ \ \ \ \ \ \ \ \ \ \ \ \ \ \ \ \ \ \ \ \ \ \ \ \ \ \ \ \ \ \ \ \ \ \ \ \ \ \ \ \ \ \ \ \ \ 

\ \ \ \ \ \ \ \ \ \ \ \ \ \ \ \ \ \ \ \ \ \ \ \ \ \ \ \ \ \ \ \ \ 

In Table 5 -- Table 7, we see that the theoratical meson spectrum agrees well
with the experiment results \cite{Meson(2000)}.\ \ \ \ \ \ \ \ \ \ \ \ \ \ \ \ \ \ \ \ \ \ \ \ \ \ \ \ \ \ \ \ \ \ \ \ \ \ \ \ \ \ \ \ \ \ \ \ \ \ \ \ \ \ \ \ \ \ \ \ \ \ \ \ \ \ \ \ \ \ \ \ \ \ \ \ \ \ \ \ \ \ \ \ \ \ \ \ \ \ \ \ \ \ \ \ \ \ \ \ \ 

\section{Predictions and Discussion}

\subsection{Some New Particles}

1. New Quarks

{\small \ \ \ \ \ \ \ \ \ \ \ \ \ \ \ \ \ \ \ \ \ \ \ \ \ \ }%

\begin{tabular}
[c]{|l|l|l|l|l|l|l|}\hline
{\small u}$^{\prime}${\small (3)} & {\small q}$_{N}^{\ast}(${\small 3091)} &
q$_{S}^{\ast}$(2551) & q$_{S}^{\ast}$(4271) & q$_{S}^{\ast}$(10031) &
q$_{b}^{\ast}$(9951) & q$_{b}^{\ast}$(15811)\\\hline
{\small d}$^{^{\prime}}$(6) & {\small q}$_{\Delta}^{\ast}(${\small 3091)} &
q$_{C}^{\ast}$(2961) & q$_{C}^{\ast}$(6591) & q$_{C}^{\ast}$(13791) &
q$_{\Omega}^{\ast}$(3071) & q$_{\Omega}^{\ast}$(3711)\\\hline
\end{tabular}

\ \ \ \ \ \ \ \ \ \ \ \ 

2. New Baryons

{\small \ \ \ \ \ \ \ \ \ \ \ \ \ \ \ \ \ }%

\begin{tabular}
[c]{|l|l|l|l|l|l|}\hline
N(3100) & $\Lambda^{0}$(2560) & $\Lambda^{0}$(4280) & $\Lambda^{0}$(10040) &
$\Lambda_{b}^{0}$(9960) & $\Lambda_{b}^{0}$(15820)\\\hline
$\Delta$(3100) & $\Lambda_{C}^{+}$(2970) & $\Lambda_{C}^{+}$(6600) &
$\Lambda_{C}^{+}$(13800) & $\Omega^{-}$(3080) & $\Omega^{-}$(3720)\\\hline
\end{tabular}

\ \ \ \ \ \ \ \ \ \ \ 

3. New Mesons

\ \ \ \ \ \ \ \ \ \ \ \ \ \ \ \ \ \ \ \ \ \ \ \ \ %

\begin{tabular}
[c]{|l|l|}\hline
q$_{N}^{\ast}$(931)$\overline{q_{S}^{\ast}\text{{\small (4280)}}}%
${\small =K(3597)} & q$_{N}^{\ast}$(931)$\overline{q_{b}^{\ast}%
\text{{\small (9951)}}}${\small \ =B(9504)}\\\hline
$\overline{q_{N}^{\ast}(931)}q_{C}^{\ast}${\small (6591)=D(5996)} &
q$_{S}^{\ast}$(1111)$\overline{q_{b}^{\ast}\text{{\small (9951)}}}%
${\small \ =B}$_{S}${\small (9659)}\\\hline
{\small q}$_{C}^{\ast}${\small (6591)}$\overline{q_{S}^{\ast}%
\text{{\small (1111)}}}${\small =D}$_{S}${\small (6151)} & {\small q}%
$_{S}^{\ast}${\small (4271)}$\overline{\text{{\small q}}_{S}^{\ast
}\text{{\small (4271)}}}${\small =}$\eta${\small (5926)}\\\hline
{\small q}$_{S}^{\ast}${\small (10031)}$\overline{q_{S}^{\ast}(10031)}%
${\small \ =}$\eta${\small (17837)} & q$_{N}^{\ast}$(931)$\overline{q_{\Delta
}^{\ast}(1291)}${\small =T(960)}\\\hline
{\small q}$_{C}^{\ast}${\small (13791)}$\overline{q_{C}^{\ast}(13791)}%
${\small \ =}$\psi${\small (25596)} & q$_{N}^{\ast}$(931)$\overline{q_{\Delta
}^{\ast}(1651)}${\small =T(1282)}\\\hline
{\small q}$_{b}^{\ast}${\small (9951)}$\overline{q_{b}^{\ast}(9951)}%
${\small \ = }$\Upsilon${\small (17805)} & q$_{N}^{\ast}$(931)$\overline
{q_{\Delta}^{\ast}(2011)}${\small =T(1603)}\\\hline
\end{tabular}

\ \ \ \ \ \ \ \ \ \ \ \ \ \ \ \ \ \ \ \ \ \ \ \ \ \ \ \ \ \ \ \ \ \ \ \ \ \ 

The discovery of any one of the above mesons will provide strong support for
the BCC Quark Model. The meson T(1603) with I = 2 has been
discovered{\small \ }\cite{Meson (I=2)} ($\chi(1600)$), although this still
needs confirmation. The quarks u$^{\prime}$(3) and d$^{\prime}$(6) have
already been discovered [u(1--5) and d(3--9)] \cite{Quark Mass(2000)}.

\subsection{Discussion}

1. Based on Dirac's sea concept of the vacuum, the BCC Quark Model assumes
that an infinite body center cubic quark (u and d) lattice is in the vacuum.
The quark lattice is the foundation of the quark spectrum, the baryon
spectrum, and the meson spectrum. Thus, it is the physical foundation of the
BCC Quark Model. It should be the physical foundation of the Quark Model also
since the Quark Model is an approximation of the BCC Quark Model. That the
theoretical baryon spectrum and the theoretical meson spectrum agree well with
the experimental results shows that the BCC quark lattice does exist in the vacuum.

2. After the discovery of super conductors, we could understand the vacuum
material. In a sense, the vacuum material (skeleton -- the BCC quark lattice)
works like a superconductor. Since the energy gaps are so large (the energy
gap of an electron is about 0.5 Mev and the energy gap of a proton is 939
Mev), under the transition temperature (much higher than the temperature at
the center of the sun),\ there are no electric or mechanical resistances to
any particle or to any physical body moving inside the vacuum material. The
vacuum material is a super superconductor.

3. The Quark Model assumes 6 flavored elementary quarks (u, d, s, c, b, and
t); there are the 6 flavored quark in the vacuum state; there are only the 6
flavored quarks in the excited states (from the vacuum) also. They have fixed
intrinsic quantum numbers (I, S, C, b, and Q) and static masses that cannot be
changed. Using 5 of the 6 flavored quarks (the t-quark with m=175 Gev is not
needed), the Quark Model explains the baryon spectrum and the meson spectrum.
It cannot deduce the mass spectrums of the baryons and the mesons and cannot
simply explain the decays of the baryons and the mesons (it needs to give up
the ``elementary'' concept of the 5 flavored quarks and the assumption that
the quarks d, s, c, and b can decay into other quarks). The BCC\ quark model
needs only 2 flavored quarks (u and d); there are only the 2 flavored quarks
in the vacuum state (the BCC quark lattice); all other quarks of the BCC Quark
Model are excited quarks. There are many different energy-band excited states
(including the quarks s, c, and b) with different intrinsic quantum numbers
(I, S, C, b, and Q) and static masses (the quark spectrum). Using the excited
quark spectrum, the BCC Quark Model can deduce the mass spectra of the baryons
and the mesons, and it can explain easily the decays of the baryons and the
mesons since the higher mass quarks are the higher energy excited states of
the same quark. (The higher energy excited states decay into lower energy
excited states and finally into the ground state is a physical law.)

4. A very important distinction between the BCC Quark Model and the Quark
Model is the model of the baryons. A baryon is composed of 3 quarks in the
Quark Model (proton=uud); but, in the BCC Quark Model, a baryon is made of an
excited quark (q$^{\ast}$) and 2 accompanying excited quarks ($u^{\prime}$ and
$d^{\prime}$). The two models give the same intrinsic quantum numbers (S,C,b,
I, and Q) but a different mass for each baryon. The baryon model of the Quark
Model is like a molecule (baryon) made by 3 atoms (quarks); the baryon model
of the BCC Quark Model is like 1 earth (q$^{\ast} $) with 2 moons ($u^{\prime
}$ and $d^{\prime}$). Which is correct?

5. The small masses (m$_{u}$ = 1 to 5 Mev, m$_{d}$ = 3 to 9 Mev) \cite{Quark
Mass(2000)} of the quarks are supported by a vast amount of experimental and
theoretic results. We believe that there are actually the quarks u(3) and d(6)
inside baryons. Although they are not mistakes, they are really too small to
build a stable baryon. Using the sum laws (\ref{SUM-LAWS}) and the quark
masses \cite{Quark Mass(2000)} of the Quark Model, we find the theoretical
masses of the most important baryons of the Quark Models. Similarly, using the
sum laws and the quark spectrum (\ref{Quark-Spectrum}), we find the masses of
the same baryons in the BCC Quark Model. We list the theoretical results of
the two models and the experimental results of the most important baryons
\cite{Baryon(2000)} as follows:

\ \ \ \ \ \ \ \ \ \ \ \ \ \ \ \ \ \ \ \ \ \
\begin{equation}%
\begin{tabular}
[c]{|l|l|l|l|l|l|l|l|l|}\hline
{\small Baryon} & {\small p({\tiny 938})} & {\small n({\tiny 940})} &
$\Lambda${\small ({\tiny 1116})} & $\Sigma^{0}${\small ({\tiny 1193})} &
$\Xi^{0}${\small ({\tiny 1315})} & $\Omega${\small ({\tiny 1672})} &
$\Lambda_{C}${\small ({\tiny 2285})} & $\Lambda_{b}${\small ({\tiny 5624}%
)}\\\hline
{\small Quark.} & {\small uud} & {\small udd} & {\small uds} & {\small uds} &
{\small uss} & {\small sss} & {\small udc} & {\small udb}\\\hline
{\small Mass } & {\small 12} & {\small 15} & {\small 132} & {\small 132} &
{\small 249} & {\small 369} & {\small 1259} & {\small 4209}\\\hline
{\small BCC } & {\small u}$^{\ast}${\small u}$^{^{\prime}}${\small d}%
$^{^{\prime}}$ & {\small d}$^{\ast}${\small u}$^{^{\prime}}${\small d}%
$^{^{\prime}}$ & {\small q}$_{S}^{\ast}${\small u}$^{^{\prime}}$%
{\small d}$^{^{\prime}}$ & {\small q}$_{\Sigma^{0}}^{\ast}${\small u}%
$^{^{\prime}}${\small d}$^{^{\prime}}$ & {\small q}$_{\Xi^{0}}^{\ast}%
${\small u}$^{^{\prime}}${\small d}$^{^{\prime}}$ & {\small q}$_{\Lambda
}^{\ast}${\small u}$^{^{\prime}}${\small d}$^{^{\prime}}$ & {\small q}%
$_{C}^{\ast}${\small u}$^{^{\prime}}${\small d}$^{^{\prime}}$ & {\small q}%
$_{b}^{\ast}${\small u}$^{^{\prime}}${\small d}$^{^{\prime}}$\\\hline
{\small Mass } & {\small 940} & {\small 940} & {\small 1120} & {\small 1210} &
{\small 1300} & {\small 1660} & {\small 2280} & {\small 5540}\\\hline
{\small Exper.} & {\small 938} & {\small 940} & {\small 1116} & {\small 1193}%
& {\small 1315} & {\small 1672} & {\small 2285} & {\small 5624}\\\hline
\end{tabular}
\label{Masses of Baryons}%
\end{equation}
Comparing the theoretical results of the Quark Model with the experimental
results (\ref{Masses of Baryons}), we find that the theoretical baryon masses
of the Quark Model are too small to match the experimental masses. Thus, there
will be another quark with a large mass inside the baryon. According to the
BCC Quark Model, this quark with a large mass really exists in a baryon -- it
is excited quark (q$^{\ast}$) (\ref{Masses of Baryons}); the quarks with small
masses are the accompanying excited quarks ($u^{\prime}$ and $d^{\prime}$)$.$
The theoretical masses of the baryons of the BCC Quark Model agree well with
the experimental results. Thus, the experimental results support the baryon
model ({\small q}$^{\ast}${\small u}$^{^{\prime}}${\small d}$^{^{\prime}})$ of
the BCC Quark Model. Since the mass of $q_{N}^{\ast}$ (931 Mev) is closest to
the mass of a proton or a neutron, the mass of the quark $q_{N}^{\ast}$(931
Mev) [u$^{\ast}$(931), d$^{\ast}$(931)] was misidentified as the mass of a
proton (938 Mev) and a neutron (940 Mev) in the experiments. Therefore, we
only find the small masses of the quarks in the Quark Model. In fact, the
small masses of u and d in the Quark Model are very important results -- they
are very strong evidence of the accompanying excited quarks ($u^{\prime}$ and
$d^{\prime}$).

6. Another important distinction between the BCC Quark Model and the Quark
Model is the model of the unflavored mesons with I = 0 ($\eta,\omega,\phi$, h,
and f). The meson model of the Quark Model is the mixture of a 3
quark-antiquark pairs (u$\overline{u}$, d$\overline{d}$, and s$\overline{s})$.
For example,
\begin{equation}%
\begin{tabular}
[c]{|l|l|}\hline
$\eta(547)=\eta_{8}\cos\theta_{p}-\eta_{1}\sin\theta_{p},$ & $\eta^{\prime
}(958)=\eta_{8}\sin\theta_{p}+\eta_{1}\cos\theta_{p},$\\\hline
where $\eta_{1}=(u\overline{u}+d\overline{d}+s\overline{s})/\sqrt{3},$ & and
$\eta_{8}=(u\overline{u}+d\overline{d}-2s\overline{s})/\sqrt{6}.$\\\hline
\end{tabular}
\label{Mixture Mesons}%
\end{equation}
The Quark Model needs a parameter $\theta_{p}$. For different mesons,
$\theta_{p}$ has the following different values: for $\eta$ and $\eta^{\prime
}$,$\theta_{p}$= -10$^{0}$; for $\phi$ and $\omega$, $\theta_{p} $ = 39$^{0}$;
for f$_{2}^{^{\prime}}$(1525) and f$_{2}$(1270), $\theta_{p}$ = 28$^{0}$; for
$\phi_{3}$(1850) and $\omega_{3}$(1670), $\theta_{p} $ = 29$^{0}$. The meson
model of the BCC Quark Model is one quark-antiquark pair. For the same mesons,
according to the BCC Quark Model, we have%

\begin{equation}%
\begin{tabular}
[c]{|l|}\hline
q$_{S}^{\ast}$(1111)$\overline{q_{S}^{\ast}(1111)}$=$\eta$(549) [$\bullet\eta
$(547)] ,\\\hline
q$_{S}^{\ast}($1391)$\overline{q_{S}^{\ast}(1391)}$=$\eta$(952)[$\bullet
\eta^{\prime}$(958)] .\\\hline
\end{tabular}
\label{One Quark-Antiquark Pair}%
\end{equation}
Comparing the two models, the meson model of the BCC Quark Model is better. It
does not need the the mixture of the 3 quark-antiquark pairs or the parameter
($\theta_{p}$).

7. Since the quarks (q$_{u}^{\ast}$, q$_{d}^{\ast}$, q$_{S}^{\ast}$,
q$_{c}^{\ast}$, q$_{b}^{\ast}$...) are the excited states of the elementary
quarks, the SU(3), SU(4), SU(5),..., SU(N) symmetry groups (flavor) based on
the quarks (q$_{u}^{\ast}$, q$_{d}^{\ast}$, q$_{S}^{\ast}$, q$_{c}^{\ast} $,
q$_{b}^{\ast}$...) are a natural extension of SU(2) based on the elementary
quarks (u and d). They do not need any assumptions. Although there are large
mass differences among these quarks, the SU(N) symmetries of the Quark Model
are still correct.

8. In the high-energy scattering cases, because the strong interactions
(color) of the quarks are short range and saturable (a 3 different color
system is a colorless system), we can only consider the 3 quark system
($q^{\ast}u^{\prime}d^{\prime}$) in a baryon case and the 2 quark system
(q$_{i}^{\ast}$ and $\overline{q_{j}^{\ast}}$) in a meson case{\small .} We do
not need to consider the whole BCC quark lattice. Thus, in high-energy
scattering cases, the Quark Model is an excellent approximation of the BCC
Quark Model.

\section{Conclusions\ \ \ }

1. There is a body center cubic quark (u, d) lattice in the vacuum. The vacuum
material (skeleton -- the BCC quark lattice) is a super superconductor with
super-high energy gaps and a super-high transition temperature. It has the
body center cubic periodic symmetries. We think that the body center cubic
periodic symmetries might be some of the ``have not yet been identified''
symmetries \cite{Wilczek2}.

2. There are only 2 (not 6) flavored elementary quarks (u and d) in the vacuum
state; other quarks (s, c, b...) can be deduced (energy band excited states).
There are always 2 accompanying excited quarks, $u^{\prime}(3)$ and
$d^{\prime}(6)$, accompanying each excited quark q$^{\ast}$. There is the
excited quark spectrum (\ref{Quark-Spectrum}) (many more than the 6 flavored
quarks) in the excited states. These excited quarks are the energy band
excited states of the elementary quarks.

3. The small masses of the quark u(1 to 5 Mev) and the quark d(3 to 9 Mev) are
very strong evidence of the 2 accompanying excited quarks [$u^{\prime}(3)$ and
$d^{\prime}(6)$]. They show that the accompanying excitation concept of the
BCC Quark Model is correct.

4. Using the quark spectrum, we deduce the baryon spectrum and the meson
spectrum. They agree well with experimental results. We can also explain the
decay of the (excited) quarks, the baryons, and the mesons since the
higher-mass quarks and the lower-mass quarks are all the excited states of the
same elementary quarks.

5. According to the intrinsic quantum numbers (S, C, b, I, and Q), the 5
elementary quarks (u, d, s, c, and b) of the Quark Model are just the 5 ground
states of the quark spectrum of the BCC\ Quark Model (\ref{Quark-Spectrum}).
The Quark Model only uses these 5 quarks to explain the baryon spectrum and
the meson spectrum. Therefore, the Quark Model is the 5-ground-state
approximation of the BCC Quark Model. We will discuss the sixth quark (t)
later since the baryon spectrum and the meson spectrum do not need it now.

6. The number of the flavored elementary quarks will change from 6 (u, d, s,
c, b, and t) to 2 (u and d). The small quark masses of the Quark Model should
be changed as follws: u(1 to 5) $\rightarrow$ u(931), d(3 to 9) $\rightarrow$
d(931), s(75 to 170) $\rightarrow$ s(1111), c(1150 to 1350)$\rightarrow
$c(2271), and b(4000 to 4400) $\rightarrow$ b(5531). The baryon model will
change from 3 excited quarks to 1 excited quark q$^{\ast}$ with 2 accompanying
excited quarks [u$^{^{\prime}}(3)$ and d$^{^{\prime}}(6$)]. The meson model of
the unflavored mesons ($\eta,\varpi,\phi$, h, and f) with\ isospin I = 0 will
change from the mixture of the 3 quark pairs (u$\overline{u}$, d$\overline{d}%
$, and s$\overline{s})$ to 1 pair of a quark and an antiquark. The confinement
concept of the Quark Model should be replaced by the accompanying excitation
concept of the BCC Quark Model. According to the accompanying excitation
concept, any excited\ (from the vacuum) quark q$^{\ast}$is always accompanied
by the 2 accompanying excited quarks $u^{\prime}$ and $d^{\prime}$. They
cannot be separated. Therefore, individually free quarks can never be seen.\ 

7. Due to the existence of the vacuum material, all observable particles are
constantly affected by the vacuum material (vacuum state quark lattice). Thus,
some laws of statistics (such as fluctuation) cannot be ignored.

8. Although the Quark Model needs modification, its principles are still
correct. The relevant principles are as follows: a baryon is composed of 3
quarks; a meson is made by a quark and an antiquark; the sum laws [the quantum
numbers of a baryon (a meson) are the sum of the same quantum numbers of the
quarks that compose the baryon (meson)]; the confinement idea (individual free
quarks can never been seen); and the decay assumption (the higher mass quarks
can decay into lower mass quarks). After the modification, the Quark Model has
a solid physical foundation. It is more reasonable, more easy to understand,
more useful (it can now deduce the mass spectrum of the baryons and the
mesons), and without a need for the parameters m$_{S}$, m$_{C}$, m$_{b}$, and
$\theta_{p}$.

9. Quantum Chromodynamics, of course, does not require a change. In the high
energy scattering cases, the Quark Model will work well and does not need a
change now.

10. The BCC Quark Model is really a good modification of the Quark Model, but
it is a phenomenological model. Furthermore, we need to consider the symmetry
wave functions to find the angular momenta and parities of the quarks, the
mesons, and the baryons (see the next paper by Xin Yu and Jiao-Lin Xu, ``The
symmetry wave functions of the Quarks in the BCC Quark Model'').

\begin{center}
\bigskip\textbf{Acknowledgment}
\end{center}

I thank Dr. Xin Yu very much for checking the calculations of the energy bands
and for helping to write this paper. I sincerely thank Professor Robert L.
Anderson for his valuable advice. I also acknowledge\textbf{\ }my indebtedness
to Professor D. P. Landau for his help. I especially thank W. K. Ge, J. S.
Xie, Y. S. Wu, H. Y. Guo, S. Chen, Z. Y. Wu, and Fugao Wang for help.


\begin{thebibliography}{99}
\bibitem{QuarkModel}M. Gell-Mann, Phys. Lett.\textbf{\ 8}, 214 (1964); G.
Zweig, CERN Preprint Th- 401, Th-412 (1964).

\bibitem{Hand in}The Oecd Frum, \textit{Particle Physics (Head of Publications
Service, OECD) 55 (1995). }

\bibitem{Marshah}R. E. Marshah, \textit{Conceptual Foundations of Modern
Particle Physics} (World Scientific, Singapore, 1993) xxi (Foreword).

\bibitem{Confinement}A. Pais, Rev. Mod. Phys., \textbf{71 No. 2,} S16 (1999).

\bibitem{Quark Mass(2000)}Particle Data Group, D. E. Groom et al., Eur. Phys.
J. C \textbf{15}, 26 (2000).

\bibitem{BCC Quark Model}J. L. Xu, hep-ph/0106340.

\bibitem{Quark Spectrum}J. L. Xu, hep-ph/0106340 (Section V. A).

\bibitem{Baryon Spectrum}J. L. Xu, hep-ph/0106340 (Section V. B).

\bibitem{Net-Meson}J. L. Xu and X. Yu, hep-ph/0210090.

\bibitem{Meson(2000)}Particle Data Group, D. E. Groom et al., Eur. Phys. J. C
\textbf{15}, 27 (2000).

\bibitem{BCC MODEL}J. L. Xu, hep-ph/0106340 (Section I).

\bibitem{Secsion II}J. L. Xu, hep-ph/0106340 (Section II).

\bibitem{Real value of S}We were enlightened by the Monte Carlo methods: K.
Binder, \textit{Monte Carlo Methods in Statistical Physics} (Springer-Verlag,
Berlin 1979); D. P. Landau, Phys. Rev. \textbf{B} \textbf{14}, 4054 (1979).

\bibitem{Energy Band}J. Callaway, \textit{Energy Band\ Theory} (Academic
Press, New York and London, 1964); H. Jones, \textit{THE THEORY OF BRILLOUIN
ZONES AND ELECTRONIC STATES IN CRYSTALS }({\small NORTH-HOLLAND PUBLISHING
COMPANY, AMSTERDAM), }117 (1975).

\bibitem{GellMann}M. Gell-Mann,\textit{\ Nuovo Cim. Supp.} \textbf{4}, 848
(1956); K. Nishijima, \textit{Prog. Theor. Phys}. \textbf{13}, 285 (1955).

\bibitem{Renormal}S. Weinberg, \textit{The Quantum Theory of Frields}
({\small CAMBRIDGE}, New York, 1995), 499 (1995). \ \ \ \ \ \ \ \ \ \ \ \ 

\bibitem{Baryon(2000)}Particle Data Group, D. E. Groom et al., Eur. Phys. J. C
\textbf{15}, 52 (2000).

\bibitem{Meson (I=2)}Particle Data Group, D. E. Groom et al., Eur. Phys. J. C
\textbf{15}, 51, 457 (2000); ARGUS coll., H. Albrecht et al, Phys. Lett.,
217B, 205 (1989); CELLO Coll., H. J. Behrend et al., Phys. Lett. 218B, 493
(1989); ARGUS coll., H. Albrecht et al, Z. Phys. C 50, 1 (1991).

\bibitem{Wilczek2}F. Wilczek, Rev. Mod. Phys., \textbf{71 No. 2}, S85 (1999). \ \ \ 
\end{thebibliography}
\end{document}